# Blockchain technologies in the design of Industrial Control Systems for Smart Cities

Gabriela Ahmadi-Assalemi[1], Haider Al-Khateeb[1*]
G.Ahmadi-Assalemi@wlv.ac.uk, H.Al-Khateeb@wlv.ac.uk
[1] Cyber Quarter – Midlands Centre for Cyber Security / University of Wolverhampton, UK
* Correspondence

**Abstract** –The proliferation of sensor technologies in Industrial Control Systems (ICS) helped to transform the environment towards better automation, process control and monitoring. However, sensor technologies expose the smart cities of the future to complex security challenges. Luckily, the sensing capabilities also create opportunities to capture various data types, which apart from operational use can add substantial value to developing mechanisms to protect ICS and critical infrastructure. We discuss Blockchain (BC), a disruptive technology with applications ranging from cryptocurrency to smart contracts and the value of integrating BC technologies into the design of ICS to support modern digital forensic readiness.

**Keywords:** *Cyber-Physical Systems, Digital Investigation, Digital Forensic Readiness, Digital Evidence, Digital Witness, Data Integrity, Supervisory Control and Data Acquisition, Distributed Ledger*

## 1. Introduction

*1.1. The Evolving ICS Threat Model*

Modern Industrial Control Systems (ICS) have emerged as critical components of smart cities supporting the operations of industrial facilities including critical infrastructures. Unlike in the past, modern ICS integrate with disruptive technologies creating complex, distributed and interdependent cyber-physical-natural ecosystems [1]. The fusion of Operational Technologies (OT) and Information Communication Technologies (ICT) increases process control, monitoring and automation. However, the integration of devices, sensors and ubiquitous connectivity converging the physical and digital realms create a substantial attack surface with numerous attack vectors. Against this backdrop, ICS are an attractive target for increasingly sophisticated and hostile threat actors, including Advanced Persistent Threats (APT). Besides targeted attacks from external adversaries, insider threats such as malicious intent, accidental hazards and professional errors are a substantial cybersecurity challenge. Additionally, ICS are predisposed to challenges resulting from the disparity between security and operational priorities of ICT and OT, which further complicates the protection of ICS from cybersecurity threats [1, 2]. Hence, ICS have to adapt to the evolving threat landscape with effective countermeasures as part of defence-in-depth capability.

*1.2. Blockchain competencies in cyber security*

Blockchain (BC) emerged from the underpinning technology behind the Bitcoin cryptocurrency [3, 4]. Leveraged for financial transaction recording, BC has unique characteristics. It is composed of cryptographically chained immutable blocks that form a trusted, shared and distributed ledger of transactions. The blocks in the BC are kept by peer-to-peer distributed management adopting consensus algorithms without needing a central authority or another intermediary [4-6]. Notably, nodes managing a BC are constrained by the use of the same consensus algorithm. BC is intrinsically incremental with each block being append-only, linked using secure hashes to the previous and subsequent blocks. The block comprises the hash, random nonce, root hash, the timestamp and the metadata of all transactions immutably recorded with the ability to trace back to the Genesis block. Genesis is the name given to the first block in a given BC. Due to the security capabilities including the decentralised architecture of trusted sources, cryptographic security, authenticity, responsibility for integrity and fault tolerance, BC has the potential to support securing the Internet of Things (IoT) [4, 5, 7]. Scientific literature suggests that leveraging BC to address the security of Cyber-Physical Systems (CPS) and critical infrastructures to support modern Digital Forensics and Incident Response (DFIR) is an active research area [6, 8, 9].

## 2. Application of Blockchain technology in ICS

*2.1. Secure Logs, Integrity and Traceability of Digital Evidence*
Logs systematically record events in digital systems that help understand ongoing and occurred events. Secure logs-protection is a well-understood technique in computer security to maintain the integrity of the logs to support incident responders. However, Digital Forensics (DF) readiness in ICS is a recent phenomenon. The need is attributed to the increasing and well-organised cyber crimes against ICS and critical infrastructures [10]. Apart from investigating digital crimes, logs can be leveraged to deal with insider threats such as accidental hazards and professional errors. Hence, secure logs form an important part of DF readiness. In a digital investigation, any piece of data is a potential Digital Evidence (DE) artefact. However, the method in which DE has handled influences whether that digital artefact is admissible in the Court of Law. For example, DE requires a Digital Chain-of-Custody (DCoC), which maintains traceability of the digital artefact to ensure attribution, specifically referencing the chronology of ownership, custody and location of the DE.

*2.2. The concept of Digital Witness in ICS*
Despite ICS initially designed as standalone systems with segregated ICT and OT realms, modern ICS are networked, complex, distributed and interdependent ecosystems exposed to external and insider threats [1]. If disrupted, ICS could result in devastating consequences with a serious societal and economic impact [1, 2, 6, 11]. Due to the dynamic and complex threat landscape, proactive cyber defence is increasingly critical in ICS. Reactive cyber defence capabilities alone are insufficient to effectively mitigate threats. DF readiness in ICS provides support for post-incident investigations, events reconstruction and threat modelling, which help identify and mitigate design and production shortfalls that threaten the operations of ICS. DF in this context is referred to as a branch of forensic science that covers established skills and capabilities in the cybersecurity industry and diverse digital technologies that can be exploited by adversaries [8].

The basic hierarchy of an ICS architecture consists of three distinct layers, as shown in Figure 1. The physical plant sensors in the physical control layer ubiquitously generate data. Apart from operational activities such as process control and automation, physical plant sensors could collaborate on providing specific cyber threat intelligence from the physical plant's sensing capabilities. The following study [10] utilised data from ICS physical plant sensors coupled with Machine Learning (ML) to profile anomalous behaviour and quantify cyber risk in an ICS environment as part of a layered security approach to increase resilience to cyber-attacks. The study focused on operational infrastructure at the physical control layer to address accidental and malicious activities. Profiling anomalous behaviour from physical sensors data using innovative ML methods creates an opportunity for incident responders to support their investigations. However, in pervasive systems particularly critical infrastructures it is not always viable to seize physical media to gather DE.

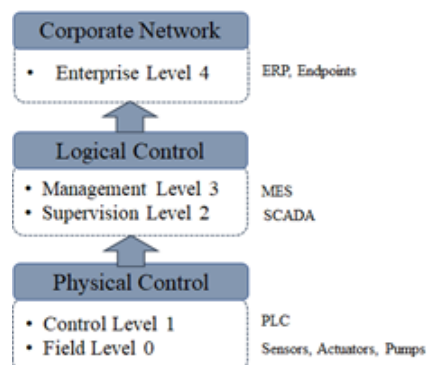

*Figure 1 Simplified ICS Architecture*

Against this backdrop, due to its security capabilities, BC technology can be integrated into ICS as an enabler to achieve modern DF readiness. While the surge of disruptive technologies integrated into ICS introduces complexities and increases their attack surface, BC presents an opportunity to leverage the concept of "Digital Witnesses" (DW) to support investigations, as illustrated in Figure 2. We refer to DW as cyber-physical objects with functional sensing capability to confirm a crime-related event [6, 8, 12], in this instance extended to accidental and malicious anomalous activities detected from ICS physical plant sensors.

The role of the DW is to identify and preserve DE artefacts that can be stored on the device or transferred to other devices in the cloud, also known as Hearsay DW. To describe the DE data lifecycle from creation to destruction, we refer to the Digital Forensic Research Workshop (DFRWS), the ISO/IEC 27050 general frameworks and adopt the categorisation of the data lifecycle context proposed by [8]. Additionally, to achieve admissibility, BC has a unique advantage to initiate and maintain a DCoC. For example, the following study [6] introduced a tracking and liability attribution framework leveraging DW to enable the tracking of objects' behaviour within smart controlled business environments to detect insider threats. The authors proposed a framework leveraging BC technology to achieve DF readiness by establishing a DCoC and introduced the concept of DW to support post-incident investigations in smart controlled work environments. Another study [12] investigated the use of DW in personal devices, where personal devices can acquire, store and transmit DE to an authorised entity reliably and securely.

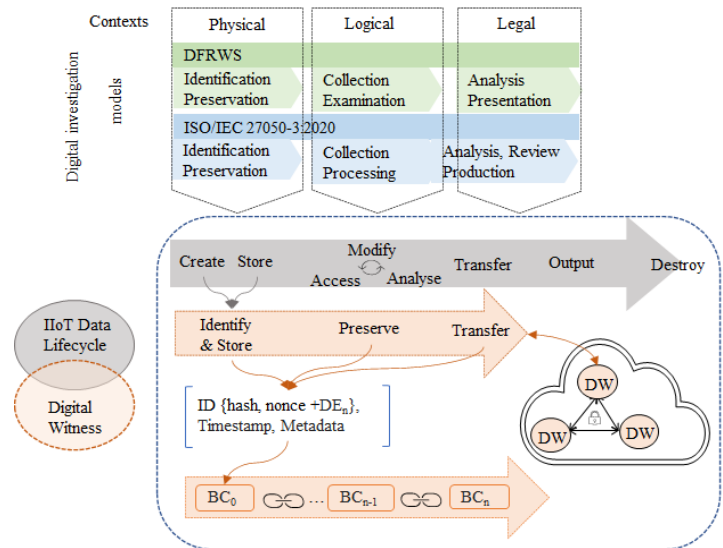

*Figure 2 Concept of Digital Witness in ICS*

## 3. Blockchain in the Design of ICS

### 3.1. Coupling of technologies

Figure 3 shows simplified elements and architecture of Industrial Automation and Control Systems (IACS) classified into the three layers introduced in Figure 1 and five distinct levels. The Physical Control Layer consists of the Field and Control Levels. The Field Level contains the field instruments and is the lowest level of the control stack hierarchy. This level includes sensors, pumps and actuators that are directly connected to the plant or equipment using the field-level networks. They generate the data that will be used by the other levels to supervise and control the processes. The Control Level uses Programmable Logic Controllers (PLC). They are adapted industrial digital computers that control the manufacturing processes. PLC link the field instruments with the **Supervisory Control and Data Acquisition (SCADA)** host software using control networks. SCADA operates at the Logical Control Layer which consists of the Supervision and Management layers interconnected by the plant network. SCADA monitors maintain and engineers processes and instruments whereas the Manufacturing Execution System (MES) is responsible for process scheduling, material handling, maintenance, and inventory. The Corporate Network Layer and the corresponding Enterprise Level is the top level of the industrial automation which manages the whole control or automation system. This level utilises Enterprise Resource Planning (ERP) systems for commercial activities including production planning, customer and market analysis, orders and sales. Furthermore, as illustrated in Figure 3, ICS are automated, they require expert engineering knowledge, real-time processing and deal with deterministic data patterns. These systems are designed for safety, reliability and availability. They have not been developed around the traditional confidentiality, integrity and availability cybersecurity pillars with a security-by-design approach [13].

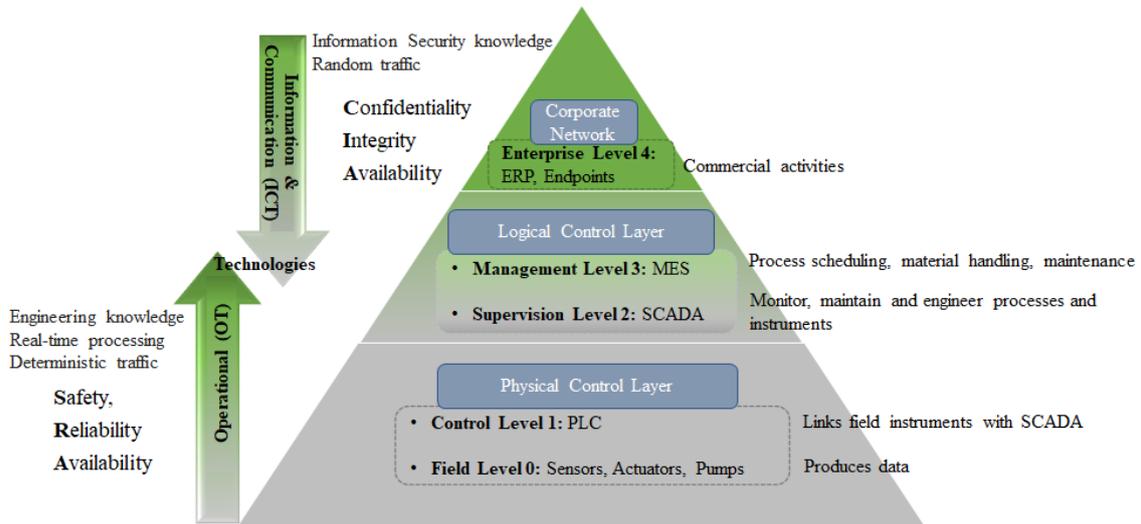

*Figure 3 Industrial Control Systems architecture and key components*

### 3.2. Integrating Blockchain into ICS Architectures

DF readiness should be carefully embedded into the ICS design and architecture as well as industrial business practice. A drawback of using BC technologies is the computational complexity of the public BC model. Despite the public BC being based on Proof-of-Work and while it can withstand up to 50% compromised nodes, the implementation of the consensus protocol is capable of fewer transactions per second. However, the Proof-of-Authority leverages pre-authorised validators suited for a permissioned network. The related private BC Practical Byzantine Fault Tolerance or Stellar Consensus is less computationally demanding, therefore functionally capable of higher throughputs. That being said, they require a higher number of trustworthy nodes. Nonetheless, a key characteristic of anomaly detection in ICS is that all objects must be known and pre-registered due to digital identity and access control. Hence, a permissioned BC is more likely to provide the throughput, manageability, traceability, trust and integrity across a complex interdependent and distributed ICS within a common framework. BC technology should be engineered and built into the ICS' design to achieve a verifiable audit trail, as shown in Figure 4, to support incident responders and facilitate digital forensic readiness as part of proactive defence-in-depth.

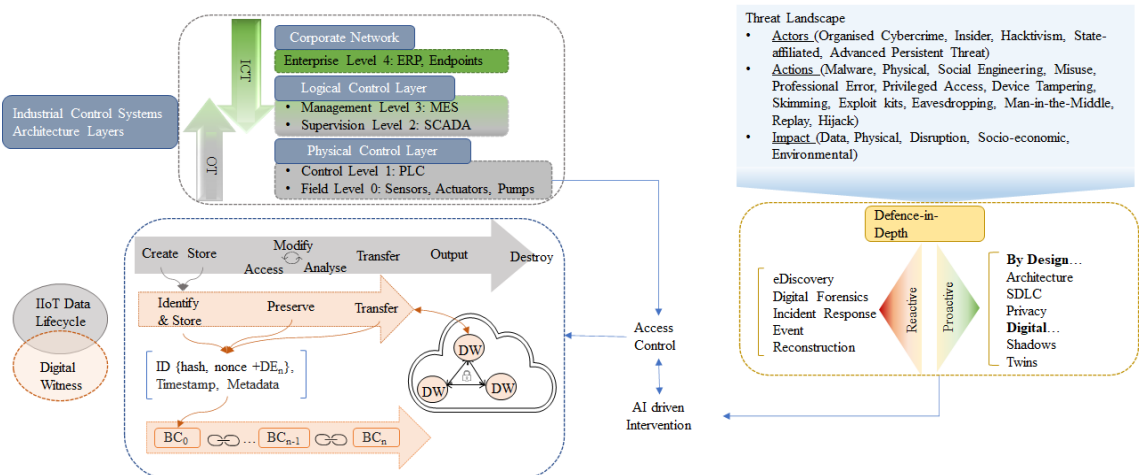

*Figure 4 Integrating Blockchain technology into ICS design*

### 3.3. Use Case: ICS Digital Forensic-enabled digital investigation

Alongside the value gained from the operational use of sensors-generated data, these data sources create multi-layered opportunities including support for developing intelligence capabilities and DF readiness. Firstly, Figure 5 demonstrates typical participants and how their interactions relate to a BC enabled ICS architecture for the physical sensors data pipeline. DWs are considered cyber-physical objects which are functionally capable to maintain admissible DE including receiving, storing and transferring DE following a pre-determined ruleset. *Employees and CPS objects* include authorised persons and smart physical or virtual objects with tamper-proof storage capable of performing tasks. *The supply chain* is considered to have a role in the development, operation and maintenance of ICT and OT concerning products, components, environmental and system parameters. *The incident investigation* includes the internal physical and virtual entities required for anomalous behaviour analysis and gatekeeping coordination of DE to legal authorities such as law enforcement agencies and the Courts of Law. Next, besides the participants and their interactions, the architectural approach requires the integration of distinct composable elements comprising of data sources, innovative use of ML techniques and BC technologies, as referenced in Figure 5. On the premise of a digital investigation, all defined data sources are potentially DE analysed by leveraging AI-driven predictive modelling. Finally, depending on the threat model, permissioned BC with smart contracts can be applied to control the ownership transfer at authorised hand-off gates to the DF investigation pipeline.

The threat landscape affecting ICS includes conventional IT and specific OT threats that range from external adversaries including Advanced Persistent Threats (APT) to the prevalence of insider threats including social challenges such as accidental hazards, social engineering and disgruntled employees. In summary, the threat model for this use case could include:

- ICS interconnectivity with public networks could result in resourceful adversaries exploiting an attack vector and gaining access to the logic and the physical control layers, see Figure 3. This could result in the alteration of values resulting in an inconsistency between the actual and expected state of the physical process resulting in anomalous behaviour captured within the physical sensors produced data.
- We could also suppose internal factors such as social engineering, disgruntled employee and supply chain exposure. Such actors have authorised unmonitored access to the operational infrastructure. This could result in alteration of the software and environmental parameters configuration, which could cause deviation from the expected data patterns. Internal threats are often underestimated and challenging to detect [6, 10].

Either use case could potentially alter the physical processes resulting in physical damage with an impact on the wider society.

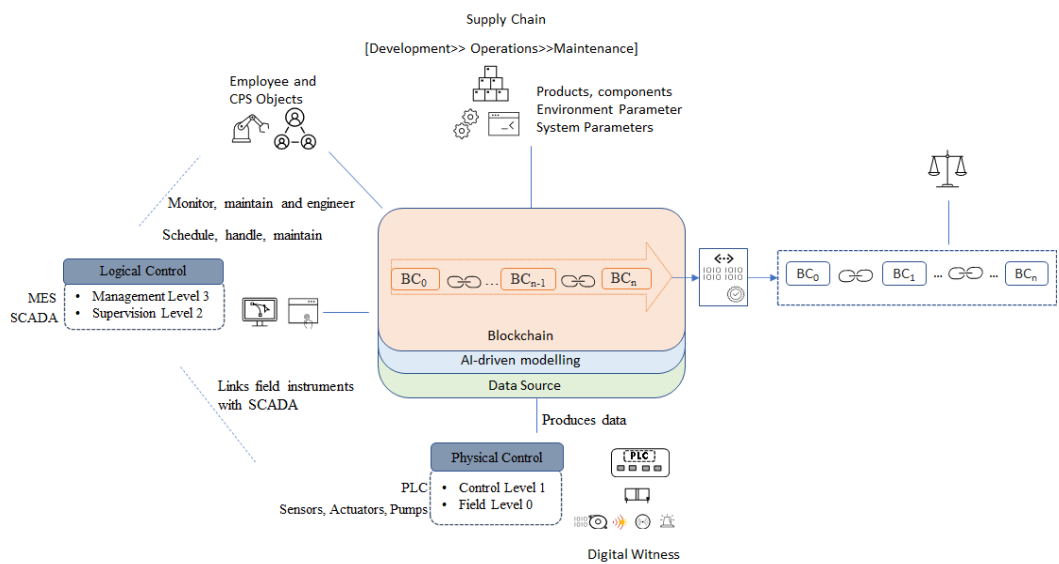

*Figure 5 BC enabled ICS general framework participants' interactions*

### 4. Regulatory Considerations

The regulatory landscape is multifaceted. Apart from the Industrial Internet of Things (IIoT) Reference Architecture [14, 15], IIoT systems have consumer-centric industry-specific standards and regulatory compliance requirements for information handling such as the Health Insurance Portability and Accountability Act (HIPPA) [15] or the General Data Protection Regulation (GDPR). Specific to the security in IACS, a catalogue of standards is published by the International Electro-technical Commission (IEC) such as the IEC 62443 covering electronic security of control systems across several industry sectors [15, 16]. Specifically, IEC62443-3-3 which relates to the details of system security requirements and security levels has been accredited by the IEC System of Conformity Assessment Schemes for Electrotechnical Equipment and Components (IECEE). The National Institute of Standards and

Technology (NIST) NISTIR7628 revision 1 standard relates to smart grid cybersecurity [17] whereas the Institute of Electrical and Electronics Engineers (IEEE) published standards on cybersecurity for intelligent electronic devices to accommodate critical infrastructure protection [18].

Additionally, NIST produced guidelines for the Network-of-Things in the NIST SP800-183 [19], and specifically a guide to ICS security in NIST SP 800-82 revision 2 [20] while mapping to further detailed security recommendations in NIST SP 800-53 revision 5 guidelines [21]. The Department of Homeland Security in the United States issued strategic principles for the security of IoT, whilst the European Union Network and Information Security (NIS) Directive guide to securely manage the connectivity between operational environments such as ICS or SCADA and the respective enterprises. In addition, the International Organisation for Standardisation (ISO) published a security catalogue covering such as ISO/IEC 30141:2018 which focuses on IoT Reference Architecture. The ISO/IEC 27001:2013 covers Security Techniques, ISO 27002:2013 is aimed at information security management practice, and the ISO/IEC 27017:2015 specifically focuses to cloud services.

Although the regulatory landscape specific to DFIR aims to establish an international baseline it is also equally diverse, covered by several standards discussed by authors in [6, 8]. For example, the ISO/IEC 27043 focuses on incident investigation principles and processes, while ISO/IEC 27037 details digital evidence acquisition. Methods for assuring suitability and adequacy of incident investigations are covered by ISO/IEC 27041, ISO/IEC 27042 provides clarity on the analysis and interpretation phases of DE. The ISO/IEC 27050-1:2019, ISO/IEC 27050-3:2020 and ISO27050-2:2018 focus on the electronic discovery of digital evidence.

Despite ongoing efforts to develop AI standards in the realm of AI-enabled systems, standards' proliferation and fragmented approach to threats make the convergence of standards challenging with currently over 80 frameworks in AI and related ethics [22, 23]. The regulations, standards and guidelines presented in this section are not exhaustive and are aimed at demonstrating the multi-disciplined complexity of the related regulatory and standards landscape.

## 5. Conclusions and final remarks

The increase in automation and interconnectedness in ICS such as energy, nuclear and water sectors widens the attack surface. We outlined the threats ranging from external adversaries to insiders creating genuine security concerns of potentially catastrophic consequences. This introduces a need to implement intelligence-driven preventative mechanisms to proactively address accidental and malicious activities.

Data produced from the physical plant sensors coupled with ML algorithms can be used to uncover unexpected behaviours in ICS. We discussed that BC is a promising emerging disruptor technology that has the potential to support DF readiness including post-incident investigations, reconstruction of events and establishing prior patterns. Besides its immutable timestamps that could be leveraged to establish DCoC, its distributed structure makes BC resilient to integrity attacks.

Furthermore, we presented a use case of how BC technology can be applied to ICS to support DF readiness. Apart from the prevalence of internal factors, resourceful adversaries will exploit new opportunities by evolving their tactics, techniques and procedures. Therefore, to mitigate threats in ICS, consideration to DF readiness is due from design, production, and operations through to decommissioning. Furthermore, organisations must understand all ICS components including data sources across the ICT and OT realms and the steps professionals can take to support modern defence-in-depth strategy and protect critical infrastructures.

In conclusion, despite BC being considered a forward-thinking disruptive technology and an enabler for innovation, the integration of BC in ICS remains limited. The direction for future research should consider how to effectively integrate BC into ICS as part of modern defence-in-depth. However, this is a complex problem that merits extensive further empirical research.

**References**

[1]     G. Li, Y. Shen, P. Zhao, X. Lu, J. Liu, Y. Liu, and S. C. Hoi, "Detecting cyberattacks in industrial control systems using online learning algorithms", Neurocomputing, vol. 364, pp. 338-348, 2019,[Online], https://doi.org/10.1016/j.neucom.2019.07.031

[2] S. Berger, O. Bürger, and M. Röglinger, "Attacks on the Industrial Internet of Things – Development of a multi-layer Taxonomy", Computers & Security, vol. 93, pp. 101790, June 2020,[Online], https://doi.org/10.1016/j.cose.2020.101790

[3] S. Nakamoto, "Bitcoin: A peer-to-peer electronic cash system", Decentralized Business Review, pp. 21260, 2008,[Online], Accessed: 17/05/2022, Available: https://www.debr.io/article/21260.pdf

[4] I. Makhdoom, M. Abolhasan, H. Abbas, and W. Ni, "Blockchain's adoption in IoT: The challenges, and a way forward", Journal of Network and Computer Applications, vol. 125, pp. 251-279, 2019,[Online], https://doi.org/10.1016/j.jnca.2018.10.019

[5] M. Banerjee, J. Lee, and K.-K. R. Choo, "A blockchain future for internet of things security: a position paper", Digital Communications and Networks, vol. 4, no. 3, pp. 149-160, 2018,[Online], https://doi.org/https://doi.org/10.1016/j.dcan.2017.10.006

[6] G. Ahmadi-Assalemi, H. M. Al-Khateeb, G. Epiphaniou, J. Cosson, H. Jahankhani, and P. Pillai, "Federated Blockchain-Based Tracking and Liability Attribution Framework for Employees and Cyber-Physical Objects in a Smart Workplace", in 2019 IEEE 12th ICGS3. London, UK, pp. 1-9, 16-18 Jan. 2019, https://doi.org/10.1109/ICGS3.2019.8688297

[7] M. A. Rahman, M. M. Rashid, M. S. Hossain, E. Hassanain, M. F. Alhamid, and M. Guizani, "Blockchain and IoT-Based Cognitive Edge Framework for Sharing Economy Services in a Smart City", IEEE Access, vol. 7, pp. 18611-18621, 2019,[Online], https://doi.org/10.1109/ACCESS.2019.2896065

[8] H. M. Al-Khateeb, G. Epiphaniou, and H. Daly, "Blockchain for Modern Digital Forensics: The Chain-of-Custody as a Distributed Ledger", Blockchain and Clinical Trial: Securing Patient Data, pp. 149-168, Cham: Springer International Publishing, 2019, https://doi.org/10.1007/978-3-030-11289-9_7

[9] G. Epiphaniou, P. Pillai, M. Bottarelli, H. Al-Khateeb, M. Hammoudesh, and C. Maple, "Electronic Regulation of Data Sharing and Processing Using Smart Ledger Technologies for Supply-Chain Security", IEEE Transactions on Engineering Management, vol. 67, no. 4, pp. 1059-1073, 2020,[Online], https://doi.org/10.1109/TEM.2020.2965991

[10] G. Ahmadi-Assalemi, H. Al-Khateeb, G. Epiphaniou, and A. Aggoun, "Super Learner Ensemble for Anomaly Detection and Cyber-risk Quantification in Industrial Control Systems", IEEE Internet of Things Journal, pp. 1-1, 2022,[Online], https://doi.org/10.1109/JIOT.2022.3144127

[11] A. Hassanzadeh, A. Rasekh, S. Galelli, M. Aghashahi, R. Taormina, A. Ostfeld, and M. K. Banks, "A review of cybersecurity incidents in the water sector", Journal of Environmental Engineering, vol. 146, no. 5, pp. 03120003, 2020,[Online], https://doi.org/10.1061/(ASCE)EE.1943-7870.0001686

[12] A. Nieto, R. Roman, and J. Lopez, "Digital Witness: Safeguarding Digital Evidence by Using Secure Architectures in Personal Devices", IEEE Network, vol. 30, no. 6, pp. 34-41, 2016,[Online], https://doi.org/10.1109/MNET.2016.1600087NM

[13] G. Ahmadi-Assalemi, H. Al-Khateeb, G. Epiphaniou, and C. Maple, "Cyber Resilience and Incident Response in Smart Cities: A Systematic Literature Review", MDPI Smart Cities, vol. 3, no. 3, pp. 894-927, Aug 2020,[Online], https://doi.org/10.3390/smartcities3030046

[14] S.-W. Lin, B. Miller, J. Durand, G. Bleakley, A. Chigani, R. Martin, B. Murphy, and M. Crawford, "The industrial internet of things volume G1: reference architecture", Industrial Internet Consortium, pp. 10-46, 2017,[Online], Accessed: 30/12/2020, Available: https://www.iiconsortium.org/IIC_PUB_G1_V1.80_2017-01-31.pdf

[15] S. Schrecker, H. Soroush, J. Molina, J. LeBlanc, F. Hirsch, M. Buchheit, A. Ginter, R. Martin, H. Banavara, and S. Eswarahally, "Industrial internet of things volume G4: security framework", Ind. Internet Consort, pp. 1-173, 2016,[Online], Accessed: 28 Dec 2020, Available: https://www.iiconsortium.org/pdf/IIC_PUB_G4_V1.00_PB.pdf

[16] B. Leander, A. Čaušević, and H. Hansson, "Applicability of the IEC 62443 standard in Industry 4.0 / IIoT", in Proceedings of the 14th International Conference on Availability, Reliability and Security, Canterbury, CA, United Kingdom, Association for Computing Machinery, pp. Article 101, 2019, https://doi.org/10.1145/3339252.3341481

[17] National Institute of Standards and Technology, "Guidelines for smart grid cyber security", 2014, https://doi.org/10.6028/NIST.IR.7628r1

[18] IEEE, "IEEE Standard for Intelligent Electronic Devices Cyber Security Capabilities", IEEE Std 1686-2013 (Revision of IEEE Std 1686-2007), pp. 1-29, 2014,[Online], https://doi.org/10.1109/IEEESTD.2014.6704702

[19] National Institute of Standards and Technology (NIST), "NIST Special Publication 800-183 Nentworks of 'Things'", Department of Commerce, USA, 2016, https://doi.org/10.6028/NIST.SP.800-183

[20] K. Stouffer, J. Falco, and K. Scarfone, "Guide to industrial control systems (ICS) security (NIST SP 800-82)", NIST special publication, vol. 800, no. 82, pp. 16-16, 2011,[Online], Accessed: 02/01/2021, Available: https://nvlpubs.nist.gov/nistpubs/SpecialPublications/NIST.SP.800-82r2.pdf

[21] National Institute of Standards and Technology Force Joint Task, "Security and Privacy Controls for Information Systems and Organizations", National Institute of Standards and Technology, 2020, https://doi.org/10.6028/NIST.SP.800-53r5

[22] World Economic Forum, "The Global Risks Report 2020", 2020,[Online], Accessed: 29 Dec 2020, Available: http://www3.weforum.org/docs/WEF_Global_Risk_Report_2020.pdf

[23] P. Cihon, "Standards for AI governance: international standards to enable global coordination in AI research & development", Future of Humanity Institute. University of Oxford, 2019,[Online], Accessed: 02/01/2021, Available: https://www.fhi.ox.ac.uk/wp-content/uploads/Standards_-FHI-Technical-Report.pdf

**Biographies**

*Gabriela Ahmadi-Assalemi* is currently pursuing a PhD degree in Cyber Security at the School of Engineering, Computing and Mathematical Sciences, University of Wolverhampton, UK. She is a Deputy Chief Information Security Officer at the University of Cambridge. Gabriela has 20 years of experience in IT technical leadership driving change in technical and strategic delivery. She has contributed research papers across journal articles and book chapters in high impact publications of international repute and presented her research at international conferences. Her research interests focus on attribution and anomalous behaviour detection in cyber-physical systems leveraging machine learning as part of security-by-design. Gabriela is a Fellow of the Higher Education Academy (FHEA) and a Member of the Institution of Engineering and Technology (MIET).

*Dr Haider al-Khateeb* holds a position as a Reader in Cybersecurity and Director of the Cybersecurity group, University of Wolverhampton. He is a Senior Fellow of the Higher Education Academy (SFHEA). His role includes leading R&D projects, consultancy, & research impact including instrumental, conceptual, and capacity building. With over 14 years in the UK's HE sector, he has delivered multi-disciplinary projects and worked with businesses to launch innovative products. Haider is actively engaged with the latest trends on Distributed Digital Forensics and Incident Response (DFIR), Cyber Resilience, Cyber Threat Intelligence (CTI), and Online Safety for vulnerable groups both at the University and Cyber Quarter – Midlands Centre for Cyber Security.